\documentclass[12pt]{article}
\usepackage{a4wide}
\usepackage{epsfig}
\usepackage{amsmath}

\newlength{\absize}
\catcode`@=11
\def\citer{\@ifnextchar
[{\@tempswatrue\@citexr}{\@tempswafalse\@citexr[]}}

\def\@citexr[#1]#2{\if@filesw\immediate
  \write\@auxout{\string\citation{#2}}\fi
  \def\@citea{}\@cite{\@for\@citeb:=#2\do
    {\@citea\def\@citea{--\penalty\@m}\@ifundefined
       {b@\@citeb}{{\bf ?}\@warning
       {Citation `\@citeb' on page \thepage \space undefined}}%
\hbox{\csname b@\@citeb\endcsname}}}{#1}} \catcode`@=12

\begin{document}
  \thispagestyle{empty}
  \pagestyle{empty}
  \renewcommand{\thefootnote}{\fnsymbol{footnote}}
\newpage\normalsize
    \pagestyle{plain}
    \setlength{\baselineskip}{4ex}\par
    \setcounter{footnote}{0}
    \renewcommand{\thefootnote}{\arabic{footnote}}
\newcommand{\preprint}[1]{%
  \begin{flushright}
    \setlength{\baselineskip}{3ex} #1
  \end{flushright}}
\renewcommand{\title}[1]{%
  \begin{center}
    \LARGE #1
  \end{center}\par}
\renewcommand{\author}[1]{%
  \vspace{2ex}
  {\Large
   \begin{center}
     \setlength{\baselineskip}{3ex} #1 \par
   \end{center}}}
\renewcommand{\thanks}[1]{\footnote{#1}}

\begin{center}
{\large \bf A Constraint between Noncommutative Parameters of
Quantum Theories in Noncommutative Space}
\end{center}
\vspace{1cm}
\begin{center}
Jian-Zu Zhang
\end{center}
\vspace{1cm}
\begin{center}
Institute for Theoretical Physics, East China University of
Science and Technology, \\
Box 316, Shanghai 200237, P. R. China
\end{center}
\vspace{1cm}
\begin{abstract}
In two-dimensional noncommutive space for the case of both
position - position and momentum - momentum noncommuting, a
constraint between noncommutative parameters is investigated. The
related topic of guaranteeing Bose - Einstein statistics in
noncommutive space in the general case are elucidated: Bose -
Einstein statistics is guaranteed by the deformed Heisenberg -
Weyl algebra itself, independent of dynamics. A special character
of a dynamical system is represented by a constraint between
noncommutative parameters. The general feature of the constraint
for any system is a direct proportionality between noncommutative
parameters with a proportional coefficient depending on
characteristic parameters of the system under study. The
constraint for a harmonic oscillator is illustrated.
\end{abstract}

\begin{flushleft}
${\ast}$ E-mail: jzzhang@ecust.edu.cn $\;$ Fax:
+86-21-64251138$\;$ Tel: +86-21-64252613
\end{flushleft}

\clearpage
\section{Introduction}
\setcounter{equation}{0}
Physics in noncommutative space \citer{CDS,DN} has been
extensively investigated in literature. This is motivated by
studies of the low energy effective theory of D-brane with a
nonzero NS - NS $B$ field background. Effects of spatial
noncommutativity are apparent only near the string scale, thus we
need to work at a level of noncommutative quantum field theory.
But based on the incomplete decoupling mechanism one expects that
quantum mechanics in noncommutative space (NCQM) may clarify some
low energy phenomenological consequences, and lead to qualitative
understanding of effects of spatial noncommutativity. In
literature NCQM  and its applications \citer{CST,Wu} have been
studied in detail. But an important issue about whether in
noncommutive space the concept of identical particles being still
meaningful and whether Bose-Einstein statistics and Fermi-Dirac
statistics being still maintained has not been resolved.

On the fundamental level of quantum field theory the annihilation
and creation operators appear in the expansion of the (free) field
operator $\Psi(\hat{x})=\int d^3k a_k(t)\Phi_k(\hat{x})+H.c.$ The
consistent multi-particle interpretation requires the usual
(anti)commutation relations among $a_k$ and $a^\dagger_k$.
Introduction of the Moyal type deformation of coordinates may
yield a deformation of the algebra between the creation and
annihilation operators. Various authors \cite{BMPV,BGMPQV} argue
important consequences for Pauli's principle in the case of
Fermi-Dirac statistics. In noncommutative quantum field theory
Poincar\'e invariance is broken and is replaced by a twisted
Poincar\'e symmetry. On the other hand this is only possible if
the statistics is equally twisted \cite{BGMPQV}. Whether the
deformed Heisenberg - Weyl algebra is consistent with Bose -
Einstein statistics is still an open issue at the level of quantum
field theory.

In this paper our study is restricted in the context of
non-relativistic quantum mechanics to elucidate this problem. We
follow the standard procedure of establishing Bose - Einstein
statistics in the ordinary quantum mechanics in commutative space,
and investigate whether Bose - Einstein statistics can be
maintained in noncommutative space. For a two dimensional
isotropic harmonic oscillator the consistency of the deformed
Heisenberg - Weyl algebra with Bose - Einstein statistics is
elucidated \cite{JZZ04a}. But this example is special. We need to
clarify the situation for general cases. We find that in the case
of both noncommutativities of position - position and momentum -
momentum Bose - Einstein statistics is guaranteed by the deformed
Heisenberg - Weyl algebra itself, independent of dynamics. The
special character of a dynamical system is represented by a
relation between noncommutative parameters. The general feature of
such a relation for any system is a direct proportionality between
noncommutative parameters with a proportional coefficient
depending on characteristic parameters of the system under study,
and such a proportional coefficient can be fixed up to a
dimensionless constant. The speciality of a two dimensional
isotropic harmonic oscillator is that the relation between
noncommutative parameters can be completely determined.

In order to demonstrate consistency between the deformed
Heisenberg - Weyl algebra and Bose - Einstein statistics, in the
following we first review the necessary background.

\section{The Deformed Heisenberg - Weyl Algebra}
\setcounter{equation}{0}
The start point is  the deformed Heisenberg - Weyl algebra. We
consider the case of both position - position noncommutativity
(space-time noncommutativity is not considered) and momentum -
momentum noncommutativity. In this case the consistent deformed
Heisenberg - Weyl algebra is \cite{JZZ04a}:
\begin{equation}
\label{Eq:xp}
[\hat x_{i},\hat x_{j}]=i\xi^2\theta\epsilon_{ij}, \qquad [\hat
p_{i},\hat p_{j}]=i\xi^2\eta\epsilon_{ij}, \qquad
[\hat x_{i},\hat p_{j}]=i\hbar\delta_{ij},\;(i,j=1,2),
\end{equation}
where $\theta$ and $\eta$ are constant parameters, independent of
the position and momentum. Here we consider the noncommutativity
of the intrinsic canonical momentum. It means that the parameter
$\eta$, like the parameter $\theta$, should be extremely small.
This is guaranteed by a direct proportionality provided by a
constraint between them (See Eq.~(\ref{Eq:cc1}) below). The
$\epsilon_{ij}$ is a two-dimensional antisymmetric unit tensor,
$\epsilon_{12}=-\epsilon_{21}=1,$ $\epsilon_{11}=\epsilon_{22}=0$.
In Eq.~(\ref{Eq:xp}) the scaling factor
$\xi=(1+\theta\eta/4\hbar^2)^{-1/2}$ is a dimensionless constant.
When $\eta=0,$ we have $\xi=1$. The deformed Heisenberg - Weyl
algebra (\ref{Eq:xp}) reduces to the one of only position -
position noncommuting.

In literature there is a tacit confusion about the difference
between the intrinsic noncommutativity of the canonical momenta
discussed here and the noncommutativity of the mechanical momenta
of a particle in an external magnetic field with a vector
potential $A_i(x_j)$ in commutative space. In the later case the
mechanical momentum is
\begin{equation}
\label{Eq:pm}
p_{mech,i}=\mu \dot x_i=p_i-\frac{q}{c}A_i,
\end{equation}
where $p_i=-i\hbar \partial_i$ is the canonical momentum in
commutative space, satisfying $[p_i,p_j]=0$. The commutator
between $p_{mech,i}$ and $p_{mech,j}$ is
\begin{equation}
\label{Eq:pm-pm}
[p_{mech,i},p_{mech,j}]=-\frac{q}{c}
\left([p_i,A_j]+[A_i,p_j]\right)= i\frac{\hbar
q}{c}\left(\partial_i A_j-\partial_j A_i\right)=i\frac{\hbar
q}{c}\epsilon_{ij3}B_3.
\end{equation}
Such a noncommutativity is determined by the external magnetic
field $\vec B$ which, unlike the noncommutative parameter $\eta$,
may not be extremely small. Thus the noncommutativity the
mechanical momenta of a particle in an external magnetic field in
commutative space is essentially different from the intrinsic
noncommutativity, the second equation in Eq.~(\ref{Eq:xp}), of the
canonical momentum in noncommutative space.

The deformed Heisenberg - Weyl algebra (\ref{Eq:xp}) can be
realizations by undeformed variables as follows (henceforth
summation convention is used)
\begin{equation}
\label{Eq:hat-x-p}
\hat x_{i}=\xi(x_{i}-\frac{1}{2\hbar}\theta\epsilon_{ij}p_{j}),
\quad
\hat p_{i}=\xi(p_{i}+\frac{1}{2\hbar}\eta\epsilon_{ij}x_{j}),
\end{equation}
where $x_{i}$ and $p_{i}$ satisfy the undeformed Heisenberg - Weyl
algebra
$[x_{i},x_{j}]=[p_{i},p_{j}]=0,\;
[x_{i},p_{j}]=i\hbar\delta_{ij}.$

It should be emphasized that for the case of both position -
position and momentum - momentum noncommuting the scaling factor
$\xi$ in Eqs.~ (\ref{Eq:xp}) and (\ref{Eq:hat-x-p}) guarantees
consistency of the framework, and plays an essential role in
dynamics. One may argues that only three parameters $\hbar$,
$\theta$ and $\eta$ can appear in three commutators (\ref{Eq:xp}),
thus $\xi$ is an additional spurious parameter and can be set to
$1.$ If one re-scales $\hat x_{i}$ and $\hat p_{i}$ so that
$\xi=1$ in Eqs.~(\ref{Eq:xp}) and (\ref{Eq:hat-x-p}), it is easy
to check that Eq.~(\ref{Eq:hat-x-p}) leads to
$[\hat x_{i},\hat
p_{j}]=i\hbar\left(1+\theta\eta/4\hbar^2\right)\delta_{ij},$
thus the Heisenberg commutation relation cannot be maintained.

\section{Consistency Between The Deformed Heisenberg - Weyl
Algebra and Bose - Einstein Statistics} \setcounter{equation}{0}
In noncommutative space the concept of identical particles being
meaningful and Bose-Einstein statistics being maintained in the
general case are elucidated by the following theorem:

{\bf Theorem} {\it In the case of both position - position and
momentum - momentum noncommuting the deformed Heisenberg - Weyl
algebra is consistent with Bose - Einstein statistics.}

Proving this theorem includes two aspects. The first aspect is to
construct the general representations of the deformed annihilation
and creation operators which satisfy the complete and closed
deformed bosonic algebra \cite{JZZ04a}. The second aspect is, by
generalizing one - particle quantum mechanics, to establish the
Fock space of identical bosons.

In the context of quantum mechanics the general representation of
the deformed annihilation and creation operator $\hat a_i$ and
$\hat a_i^\dagger$ by $\hat x_i$ and $\hat p_i$ is
\begin{equation}
\label{Eq:hat-a}
\hat a_i=c_1(\hat x_i+ic_2\hat p_i),\;
\hat a_i^\dagger=c_1(\hat x_i-ic_2\hat p_i),
\end{equation}
where $c_1$ and $c_2$ are constants and may depend on
characteristic parameters, the mass $\mu$, the frequency $\omega$
etc., of the system under study. $c_1$ and $c_2$ can be fixed as
follows. Operators $\hat a_i$ and $\hat a_i^\dagger$ should
satisfy the bosonic commutation relations $[\hat a_1,\hat
a_1^\dagger]=[\hat a_2,\hat a_2^\dagger]=1$ (to keep the physical
meaning of $\hat a_i$ and $\hat a_i^\dagger$). From this
requirement and the deformed Heisenberg - Weyl algebra
(\ref{Eq:xp}) it follows that
\begin{equation}
\label{Eq:c1-c2}
c_1=\sqrt{1/2\hbar c_2}.
\end{equation}

Following the standard procedure in quantum mechanics, starting
from a system with one particle, the state vector space of a
many-particle system can be constructed by generalizing one -
particle formulism. Then Bose - Einstein statistics for a
identical - boson system can be developed in the standard way.
Bose - Einstein statistics should be maintained at the deformed
level described by $\hat a_i$, thus operators $\hat a_i$ and $\hat
a_j$ should be commuting: $[\hat a_i,\hat a_j]=0$. From this
equation and the deformed Heisenberg - Weyl algebra (\ref{Eq:xp})
it follows that $ic_1^2\xi^2\epsilon_{ij}(\theta-c_2^2\eta)=0$.
Thus {\it the condition of guaranteeing Bose - Einstein
statistics} reads
\begin{equation}
\label{Eq:c-2}
c_2=\sqrt{\frac{\theta}{\eta}}.
\end{equation}
From Eqs.~(\ref{Eq:hat-a}), (\ref{Eq:c1-c2}) and (\ref{Eq:c-2}) we
obtain the following deformed annihilation and creation operators
$\hat a_i$ and $\hat a_i^\dagger$:
\begin{equation}
\label{Eq:aa+1}
\hat a_i=\sqrt{\frac{1}{2\hbar}\sqrt{\frac{\eta}{\theta}}}\left
(\hat x_i +i\sqrt{\frac{\theta}{\eta}}\hat p_i\right),
\hat
a_i^\dagger=\sqrt{\frac{1}{2\hbar}\sqrt{\frac{\eta}{\theta}}}\left
(\hat x_i-i\sqrt{\frac{\theta}{\eta}}\hat p_i\right),
\end{equation}
From Eqs.~(\ref{Eq:xp}) and (\ref{Eq:aa+1}) it follows that the
deformed bosonic algebra of $\hat a_i$ and $\hat a_j^\dagger$
reads \cite{JZZ04a}
\begin{equation}
\label{Eq:[a,a+]1}
[\hat a_i,\hat a_j^\dagger]=\delta_{ij}
+\frac{i}{\hbar}\xi^2\sqrt{\theta\eta}\;\epsilon_{ij},\;
[\hat a_i,\hat a_j]=0,\;(i,j=1,2).
\end{equation}
In Eqs.~(\ref{Eq:[a,a+]1}) the three equations $[\hat a_1,\hat
a_1^\dagger]=[\hat a_2,\hat a_2^\dagger]=1,\;[\hat a_1,\hat
a_2]=0$ are the same as the undeformed bosonic algebra in
commutative space; The equation
\begin{equation}
\label{Eq:[a,a+]2}
[\hat a_1,\hat a_2^\dagger] =\frac{i}{\hbar}\xi^2\sqrt{\theta\eta}
\end{equation}
is a new type. Eqs.~(\ref{Eq:[a,a+]1}) constitute a complete and
closed deformed bosonic algebra. Because of noncommutativity of
space, different degrees of freedom are correlated at the level of
the deformed Heisenberg - Weyl algebra (1); Eq.~(\ref{Eq:[a,a+]2})
represents such correlations at the level of the deformed
annihilation and creation operators.

Now we consider the second aspect. Following the standard
procedure of constructing the Fock space of many - particle
systems in commutative space, we shall take
Eqs.~(\ref{Eq:[a,a+]1}) as the {\it definition relations} for the
complete and closed deformed bosonic algebra without making
further reference to its $\hat x_i$, $\hat p_i$ representations,
generalize it to many - particle systems and find a basis of the
Fock space.

We introduce the following auxiliary operators, the tilde
annihilation and creation operators
\begin{equation}
\label{Eq:tilde-a}
\tilde a_1=\frac{1}{\sqrt{2\alpha_1}} \left(\hat a_1+i\hat
a_2\right),\;
\tilde a_2=\frac{1}{\sqrt{2\alpha_2}} \left(\hat a_1-i\hat
a_2\right),
\end{equation}
where $\alpha_{1,2}=1\pm \xi^2\sqrt{\theta\eta}/\hbar$.
From Eqs.~(\ref{Eq:[a,a+]1}) it follows that the commutation
relations of $\tilde a_i$ and $\tilde a_j^\dagger$ read
\begin{equation}
\label{Eq:tilde[a,a+]}
\left[\tilde a_i,\tilde a_j^\dagger\right]=\delta_{ij},\;
\left[\tilde a_i,\tilde a_j\right]=\left[\tilde a_i^\dagger,\tilde
a_j^\dagger\right]=0,\;(i,j=1,2).
\end{equation}
Thus $\tilde a_i$ and $\tilde a_i^\dagger$ are explained as the
deformed annihilation and creation operators in the tilde system.
The tilde number operators $\tilde N_1=\tilde a_1^\dagger\tilde
a_1$ and $\tilde N_2=\tilde a_2^\dagger\tilde a_2$ commute each
other, $[\tilde N_1,\tilde N_2]= 0.$  A general tilde state
\begin{equation}
\label{Eq:tilde-state}
\widetilde {|m,n\rangle}\equiv (m!n!)^{-1/2}(\tilde
a_1^\dagger)^m(\tilde a_2^\dagger)^n\widetilde {|0,0\rangle},
\end{equation}
where the vacuum state $\widetilde {|0,0\rangle}$ in the tilde
system is defined as $\tilde a_i\widetilde
{|0,0\rangle}=0\;(i=1,2),$ is the common eigenstate of $\tilde
N_1$ and $\tilde N_2$:
$\tilde N_1\widetilde {|m,n\rangle}=m\widetilde {|m,n\rangle}$,
$\tilde N_2\widetilde {|m,n\rangle}=n\widetilde {|m,n\rangle}$,
$(m, n=0, 1, 2,\cdots)$,
and satisfies $\widetilde {\langle m^{\prime},n^{\prime}}
\widetilde {|m,n\rangle}=
\delta_{m^{\prime}m}\delta_{n^{\prime}n}$. Thus $\{\widetilde
{|m,n\rangle}\}$ constitute an orthogonal normalized complete
basis of the tilde Fock space. In the tilde Fock space all
calculations are the same as the case in commutative space, thus
the concept of identical particles is maintained and the formalism
of the deformed Bosonic symmetry which restricts the states under
permutations of identical particles in multi - boson systems can
be similarly developed.

The theorem is proved.

It should be emphasized that in the case of both position -
position and momentum - momentum noncommuting the special feature
is when $[\hat a_{i},\hat a_{j}]=[\hat a_{i}^\dagger,\hat
a_{j}^\dagger]=0$ are satisfied, Bose - Einstein statistics is not
guaranteed. The reason is as follows. Because the new type
(\ref{Eq:[a,a+]2}) of bosonic commutation relations correlates
different degrees of freedom, the number operators $\hat N_1=\hat
a_1^\dagger\hat a_1$ and $\hat N_2=\hat a_2^\dagger\hat a_2$ do
not commute, $[\hat N_1, \hat N_2]\ne 0.$ They have not common
eigenstates. The vacuum state of the hat system is defined as
$\hat a_i|0,0\rangle=0,\;(i=1,2)$. A general hat state $\widehat
{|m,n\rangle}$ is defined as
\begin{equation*}
\widehat {|m,n\rangle}\equiv c(\hat a_1^\dagger)^m(\hat
a_2^\dagger)^n|0,0\rangle
\end{equation*}
where $c$ is the normalization constant, these states $\widehat
{|m,n\rangle}$ are not the eigenstate of $\hat N_1$ and $\hat
N_2$:
\begin{equation*}
\hat N_1\widehat {|m,n\rangle} =m\widehat
{|m,n\rangle}+\frac{i}{\hbar}m\xi^2 \sqrt{\theta\eta}\widehat
{|m+1,n-1\rangle},\; \nonumber
\end{equation*}
\begin{equation*}
\hat N_2\widehat {|m,n\rangle} =n\widehat
{|m,n\rangle}+\frac{i}{\hbar}n\xi^2 \sqrt{\theta\eta}\widehat
{|m-1,n+1\rangle}. \nonumber
\end{equation*}
Because of Eq.~(\ref{Eq:[a,a+]2}), in calculations of the above
equations we should take care of the ordering of $a_i$ and
$a_j^\dagger$ for even $i \ne j$ in the state $\widehat
{|m,n\rangle}$. The states $\widehat {|m,n\rangle}$ are not
orthogonal each other. For example, the inner product between
$\widehat {|1,0\rangle}$ and $\widehat {|0,1\rangle}$ is
\begin{equation*}
\label{Eq:1-2b} \widehat {\langle 1,0|}\widehat {
1,0\rangle}=-\frac{i}{\hbar}\xi^2 \sqrt{\theta\eta}. \nonumber
\end{equation*}
Thus $\{\widehat {|m,n\rangle}\}$ do not constitute an orthogonal
complete basis of the Fock space of a identical - boson system.

Now we investigate two issues related to this theorem: the tilde
phase space and the constraint between noncommutative parameters.

\section{The Tilde Phase Space}
\setcounter{equation}{0}
First we consider tilde phase space variables. Using
Eqs.~(\ref{Eq:aa+1}) and (\ref{Eq:tilde-a}) we rewrite $\tilde
a_i$ as
\begin{eqnarray}
\label{Eq:tilde-aa+1}
\sqrt{\alpha_1}\; \tilde a_1 =\left
(\frac{\eta}{4\theta\hbar^2}\right)^{1/4}\left (\tilde x
+i\sqrt{\frac{\theta}{\eta}}\;\tilde p^\dagger\right), \;
\nonumber\\
\sqrt{\alpha_2}\; \tilde a_2 =\left
(\frac{\eta}{4\theta\hbar^2}\right)^{1/4}\left (\tilde x^\dagger
+i\sqrt{\frac{\theta}{\eta}}\;\tilde p\right).
\end{eqnarray}
Where the tilde coordinate and momentum $(\tilde x, \tilde p)$ are
related to $(\hat x, \hat p)$ by
\begin{equation}
\label{Eq:tilde-x,p}
\tilde x=\frac{1}{\sqrt{2}}\left(\hat x_1 +i\hat x_2\right),\;
\tilde p=\frac{1}{\sqrt{2}}\left (\hat p_1-i\hat p_2\right).
\end{equation}
The tilde phase variables $(\tilde x, \tilde p)$ satisfy the
following commutation relations:
\begin{equation}
\label{Eq:tilde-[x,p]}
[\tilde x,\tilde x^\dagger]=\xi^2\theta, \;
[\tilde p,\tilde p^\dagger]= -\xi^2\eta, \;
[\tilde x,\tilde p]=[\tilde x^\dagger,\tilde p^\dagger]=i\hbar, \;
[\tilde x,\tilde p^\dagger]=[\tilde x^\dagger,\tilde p]=0.
\end{equation}
A Hamiltonian
$\hat H(\hat x,\hat p)=\hat p_i\hat p_i/2\mu +V(\hat x_i)$
with potential $V(\hat x_i)$ in the hat system is rewritten as
\begin{equation}
\label{Eq:tilde-H}
\hat H(\hat x,\hat p)= \tilde H(\tilde x, \tilde x^\dagger, \tilde
p, \tilde p^\dagger) =\left (\tilde p\tilde p^\dagger+ \tilde
p^\dagger\tilde p\right)/2\mu+\tilde V(\tilde x, \tilde x^\dagger)
\end{equation}
in the tilde system.

In some cases calculations in the tilde system are simpler than
ones in the hat system. For example, in the hat system the
Hamiltonian of a two-dimensional isotropic harmonic oscillator is
$\hat H(\hat x,\hat p)= \hat p_i\hat p_i/2\mu + \mu\omega^2 \hat
x_i\hat x_i/2$.
In the tilde system it is rewritten as
\begin{equation}
\label{Eq:tilde-H1}
\tilde H(\tilde x, \tilde x^\dagger, \tilde p,
\tilde p^\dagger)=
\left (\tilde p\tilde p^\dagger+\tilde p^\dagger\tilde
p\right)/2\mu+ \mu\omega^2\left (\tilde x\tilde x^\dagger+ \tilde
x^\dagger\tilde x\right)/2 =\hbar\left (\tilde \omega_i \tilde
N_i+\omega\right),
\end{equation}
where
$\tilde \omega_{1,2}=\alpha_{1,2}\; \omega$
are effective frequencies, the tilde number operators $\tilde N_1$
and $\tilde N_2$ have eigenvalues $n_1, n_2=0, 1, 2,\cdots$. From
Eq.~(\ref{Eq:tilde-H1}) it follows that the energy eigenvalues of
$\tilde H(\tilde x, \tilde x^\dagger, \tilde p, \tilde p^\dagger)$
are
\begin{equation}
\label{Eq:tilde-E1}
\tilde E_{n_1,n_2} =\hbar\left (\tilde
\omega_i n_i+\omega \right)=\hbar\omega \left (n_1+n_2+1
\right)+\hbar\omega\sqrt{\theta\eta}\left (n_1- n_2 \right).
\end{equation}
The last term represents the shift of the energy level originated
from effects of spacial noncommutativity. There is no shift for
zero-point energy $\omega$. It is worth noting that
Eq.~(\ref{Eq:tilde-E1}) gives the {\it exact} (non-perturbational)
eigenvalues.

Ref.~\cite{JLR} also investigated the structure of a
noncommutative Fock space and obtained eigenvectors of several
pairs of commuting hermitian operators which can serve as basis
vectors in the noncommutative Fock space. Calculations in such a
noncommutative Fock space are much complex than the above
(commutative) tilde Fock space.

\section{The Constraint Between Noncommutative Parameters}
\setcounter{equation}{0}
The structure of the deformed annihilation and creation operators
$\hat a_{i}$ and $\hat a_{i}^\dagger$ in Eqs.~(\ref{Eq:aa+1}) are
determined by the deformed Heisenberg - Weyl algebra
(\ref{Eq:xp}), independent of dynamics. The special character of a
dynamical system is encoded in the dependence of the factor
$\sqrt{\theta/\eta}$ on characteristic parameters of the system
under study. This put a constraint between $\theta$ and $\eta$
which can be determined as follows.

The general representation of the undeformed annihilation operator
$a_i$ by $x_i$ and $p_i$ is
$a_i=c_1^{\prime}(x_i +ic_2^{\prime} p_i),$
where the constants $c_1^{\prime}$ can be fixed as follows.
Operators $a_i$ and $a_i^\dagger$ should satisfy bosonic
commutation relations $[a_1,a_1^\dagger]=[a_2,a_2^\dagger]=1$.
From this requirement the undeformed Heisenberg - Weyl algebra
leads to $c_1^{\prime}=\sqrt{1/2\hbar c_2^{\prime}}$. The
undeformed bosonic commutation relation $[a_{i},a_{j}]=0$ is
automatically satisfied, so $c_2^{\prime}$ is a free parameter.
Thus the general representation of the undeformed annihilation
operator reads
\begin{equation}
\label{Eq:aa+2}
a_i=\frac{1}{\sqrt{2\hbar c_2^{\prime}}}(x_i +ic_2^{\prime}p_i),
\end{equation}
operators $a_i$ and $a_i^\dagger$ satisfy the undeformed bosonic
algebra
$[a_{i},a_{j}]=[a_i^\dagger,a_j^\dagger]=0, \;
[a_{i},a^{\dagger}_{j}]=i\delta_{ij}.$

From Eqs.~(\ref{Eq:xp}), (\ref{Eq:hat-x-p}), (\ref{Eq:hat-a}),
(\ref{Eq:c-2}) and (\ref{Eq:aa+2}) it follows that $\hat a_{i}$
can be represented by $a_{i}$ as follows:
\begin{equation}
\label{Eq:aa+3}
\hat
a_{i}=\xi(a_{i}+\frac{i}{2\hbar}\sqrt{\theta\eta}\epsilon_{ij}a_{j}),
\end{equation}

Similar to Eqs.~(\ref{Eq:xp}) and (\ref{Eq:hat-x-p}), it should be
emphasized that for the case of both position - position and
momentum - momentum noncommuting the scaling factor $\xi$ in
Eq.~(\ref{Eq:aa+3}) guarantees consistency of the framework.
Specially, it maintains the bosonic commutation relation $[\hat
a_{i},\hat a^{\dagger}_{j}]=i\delta_{ij}.$

In the limit $\theta,\eta\to 0$, the deformed operators $\hat
x_{i}, \hat p_{i}, \hat a_{i}$ reduce to the undeformed ones
$x_{i}, p_{i}, a_{i}$. Eq.~(\ref{Eq:c-2}) indicates that in this
limit $\theta/\eta$ should  keep finite. From Eqs.~(\ref{Eq:xp}),
(\ref{Eq:hat-x-p}), (\ref{Eq:hat-a}), (\ref{Eq:c-2}),
(\ref{Eq:aa+1}), (\ref{Eq:aa+2}) and (\ref{Eq:aa+3}), it follows
that
\begin{equation}
\label{Eq:c-c^prime}
c_1=c_1^{\prime},\;c_2=c_2^{\prime}.
\end{equation}
From Eqs.~(\ref{Eq:c-2}) and (\ref{Eq:c-c^prime}) we obtain the
following constrained condition
\footnote {\; Eq.~(\ref{Eq:aa+1}) is the most general
representation of the physical annihilation operator $\hat a_i$ in
noncommutative space. In literature there is an extensively tacit
understanding about the definition of the physical annihilation
operator such that {\it`` \,it is possible to construct an
infinity of the creation/annihilation operators which satisfy
exactly the bosonic commutation relations, but do not require any
constraint on the parameters such as Eq.~(\ref{Eq:cc1})"}. For
example, similar to the Landau creation and annihilation operators
(acting within or across Landau levels) involve mixing of spatial
directions in an external magnetic field, we may define the
following annihilation operator
$$
\hat {a_i^{\prime}} = \frac{\nu^{-1}}{\sqrt{2 \hbar c_2^{\prime}}}
\left[ \left( \delta_{ij} - \frac{i c_2^{\prime} \eta}{2 \hbar}
\epsilon_{ij} \right) \hat x_j + i \left(c_2^{\prime} \delta_{ij}
- \frac{i \theta}{2 \hbar} \epsilon_{ij} \right) \hat p_j
\right],$$
where $\nu = \xi (1 - \theta\eta/4 \hbar^2)$. These operators
automatically satisfy the bosonic commutation relations
$[\hat {a_i^{\prime}} , \hat {a_j^{\prime}}^{\dagger} ] =
\delta_{ij}, \hspace{0.5 cm}
[\hat {a_i^{\prime}} , \hat {a_j^{\prime}}] = [\hat
{a_i^{\prime}}^{\dagger} , \hat {a_j^{\prime}}^{\dagger}]=0. $
Moreover no constraint on the parameters $\theta$ and $\eta$ is
required apart from the obvious one $\eta \theta \ne 4 \hbar^2$.

The previous construction also indicates that it is not compulsory
to consider both position and momentum noncommutativity. Indeed,
if we take $\eta =0$, $\nu = \xi =1$ in the previous expression
for the creation/annihilation operators, we get:
$$
\hat {a_i^{\prime\prime}} = \frac{1}{\sqrt{2 \hbar c_2'}} \left[
\hat x_i + i \left(c_2' \delta_{ij} - \frac{i \theta}{2 \hbar}
\epsilon_{ij} \right) \hat p_j \right],$$ This is also perfectly
consistent.

In order to clarify the meaning of $\hat {a_i^{\prime}}$ we insert
Eqs.~(\ref{Eq:hat-x-p}) into it. It follows that
$[( \delta_{ij} -i c_2^{\prime} \eta \epsilon_{ij}/2 \hbar) \hat
x_j + i(c_2^{\prime} \delta_{ij} -i \theta \epsilon_{ij}/2 \hbar)
\hat p_j]=\xi (1 - \theta \eta/ 4 \hbar^2)(x_i
+ic_2^{\prime}p_i),$
thus
$\hat {a_i^{\prime}}=(x_i +ic_2^{\prime}p_i)/\sqrt{2 \hbar c_2'},$
which elucidates that $\hat {a_i^{\prime}}$ is just the undeformed
annihilation operator $a_i$ in Eq.~(\ref{Eq:aa+2}), not the
annihilation operator in noncommutative space. This explains that
$\hat {a_i^{\prime}}$ and $\hat {a_i^{\prime}}^{\dagger}$
automatically satisfy the undeformed bosonic commutation
relations, and no constraint on the parameters $\theta$ and $\eta$
is required.

For the case $\eta = 0$, $\nu = \xi =1$, inserting
Eqs.~(\ref{Eq:hat-x-p}) into $\hat {a_i^{\prime\prime}},$ we
obtain
$$
\hat {a_i^{\prime\prime}} = \frac{1}{\sqrt{2 \hbar
c_2^{\prime}}}(x_i +ic_2^{\prime}p_i),$$
which is the annihilation operator in commutative space again.}

\begin{equation}
\label{Eq:cc1}
\eta=K\theta,
\end{equation}
where the coefficient $K={c_2^{\prime}}^{-2}$ is a constant with a
dimension $(mass/time)^2$. Eq.~(\ref{Eq:cc1}) shows that the
general feature of such a constraint for any system is a direct
proportionality between noncommutative parameters $\eta$ and
$\theta$.

In Eq.~(\ref{Eq:cc1}) the proportional coefficient $K$ is not
determined. In the context of quantum mechanics for simple cases
the dimensional analysis can determine $c_2^{\prime}$ up to a
dimensionless constant.

As an example, we consider a harmonic oscillator. The dimension of
$c_2^{\prime}$ in Eqs.~(\ref{Eq:aa+2}) is $time/mass$. The
characteristic parameters in the Hamiltonian of a harmonic
oscillator are the mass $\mu$, frequency $\omega$ and $\hbar$. The
unique product of $\mu^{t_1}$, $\omega^{t_2}$ and $\hbar^{t_3}$
possessing the dimension $time/mass$ is $\mu^{-1}\omega^{-1}$. So
one obtains $c_2^{\prime}=\gamma/\mu\omega$, where $\gamma$ is a
dimensionless constant and can be determined as follows.

The position $x_{i}$ and momentum $p_{i}$ are, respectively,
represented by $a_i$ and $a_i^\dagger$ as
$$x_{i}=\sqrt{\frac{\gamma\hbar}{2\mu\omega}}\left(a_{i}+
a_i^\dagger\right),\;
p_{i}=-i\sqrt{\frac{\hbar\mu\omega}{2\gamma}}\left(a_{i}-
a_i^\dagger\right).$$
In the vacuum state $|0>$ the expectations of the kinetic and the
potential energy, respectively, read
\begin{equation}
\label{Eq:Ek,Ep}
\overline {E_k}=<0|\frac{1}{2\mu}p_{i}^2|0>
=\frac{\hbar\omega}{4\gamma},\;
\overline {E_p}=<0|\frac{1}{2}\mu\omega^2x_{i}^2|0>
=\frac{\gamma\hbar\omega}{4}.
\end{equation}
The special character of a harmonic oscillator is that in any
state the expectation of the kinetic energy equals to the one of
the potential energy. The condition of $\overline {E_k}=\overline
{E_p}$ leads to $\gamma=\pm 1$. Because of $\overline {E_k}\ge 0$,
the only solution is $\gamma=1$. Thus the constraint between
$\theta$ and $\eta$ for  a harmonic oscillator reads
\begin{equation}
\label{Eq:cc3}
\eta=\mu^2\omega^2\theta.
\end{equation}

The method of determining such a dimensionless constant for a
harmonic oscillator, $\overline {E_k}=\overline {E_p}$, cannot be
applied to general cases. A complete determination of the
proportional coefficient $K$ in (\ref{Eq:cc1}) based on
fundamental principles for general cases is worth elucidating in
further studies.

\section{Discussions}
\setcounter{equation}{0}
We clarify the following two points to conclude the paper.
(i) Bose - Einstein statistics can be investigated at two levels:
the fundamental level of quantum field theory and the level of
quantum mechanics. Whether the deformed Heisenberg - Weyl algebra
is consistent with Bose - Einstein statistics is still an open
issue at the level of quantum field theory. Following the standard
procedure of investigating Bose - Einstein statistics in quantum
mechanics discussions restricted at the level of quantum mechanics
are allowed and meaningful. At short distances, where spatial
noncommutativity might be relevant, one also expects quantum
mechanics to break down and to be replaced by noncommutative
quantum field theory. But studies at the level of noncommutative
quantum mechanics may explores some qualitative features of
spatial noncommutativity, and some results may survive at the
level of noncommutative quantum field theory. It is therefore
hoped that studies at the level of noncommutative quantum
mechanics may give some clue for further development.
(ii) The constrained condition (\ref{Eq:cc1}) is fixed by the most
fundamental requirement, thus can apply to any dynamical system.
Ordinary quantum mechanics is a most successful theory which has
been fully confirmed by experiments. If NCQM is a realistic
physics, possible modifications from NCQM to ordinary quantum
mechanics should be extremely small. It means that both
noncommutative parameters $\theta$ and $\eta$ should be extremely
small. This is guaranteed by Eq.~(\ref{Eq:cc1}). Furthermore, it
is understood that noncommutativity between positions is
fundamental and the parameter $\theta$ keeps the same for all
systems. Noncommutativity between momenta arises naturally as a
consequence of noncommutativity between coordinates, as momenta
are defined to be the partial derivatives of the action with
respect to the noncommutative coordinates \cite{SGT}. This means
that noncommutativity between momenta depends on dynamics. Thus
$\eta$ and the proportional coefficient $K$ between $\eta$ and
$\theta$ may depend on characteristic parameters of the
Hamiltonian (or the action) of the system under study. In simple
cases when dimensional analysis works, it can determine $K$ up to
a dimensionless constant. In order to completely fix $K$
considerations from dynamics may be necessary.

\vspace{0.4cm}

ACKNOWLEDGMENTS

\vspace{0.4cm}

This work has been supported by the Natural Science Foundation of
China under the grant number 10575037 and by the Shanghai
Education Development Foundation.

\clearpage


\begin{thebibliography}{99}
\bibitem{CDS}
A. Connes, M. R. Douglas, A. Schwarz,
JHEP {\bf 9802} (1998) 003.

\bibitem{SW}
N. Seiberg and E. Witten,
JHEP {\bf 9909} (1999) 032.

\bibitem{DN}
M. R. Douglas, N. A. Nekrasov,
Rev. Mod. Phys. {\bf 73} (2001) 977
and
references there in.

\bibitem{CST}
M. Chaichian, M. M. Sheikh-Jabbari, A. Tureanu,
Phys. Rev. Lett. {\bf 86} (2001) 2716.

\bibitem{GLR}
J. Gamboa, M. Loewe, J. C. Rojas,
Phys. Rev. {\bf D64} (2001) 067901.

\bibitem{NP}
V. P. Nair, A. P. Polychronakos,
Phys. Lett. {\bf B505} (2001) 267.

\bibitem{KD}
D. Kochan, M. Demetrian,
{\bf hep-th/0102050}

\bibitem{HK}
P-M. Ho, H-C. Kao,
Phys. Rev. Lett. {\bf 88} (2002) 151602.



\bibitem{JZZ04a}
Jian-zu Zhang,
Phys. Lett. {\bf B584} (2004) 204;\;
Phys. Rev. Lett. {\bf 93}2 (2004) 04300;\;
\;
Phys. Lett. {\bf B597} (2004) 362;\;
Qi-Jun Yin, Jian-zu Zhang,
Phys. Lett. {\bf B613} (2005) 91;\;
Jian-zu Zhang,
Phys. Lett. {\bf B639} (2006) 403;\;
Phys. Rev. {\bf D74} (2006) 124005.

\bibitem{JLR} 
S. C. Jin, Q. Y. Liu, T. N. Ruan,
{\bf hep-ph/0505048}.

\bibitem{LD}
Kang LI, Sayipjamal Dulat,
Eur. Phy. J. {\bf C46} (2006) 825.

\bibitem{WLFXY}
H. Wei, J.-H Li, R.-R. Fang, X.-T. Xie, X.-X. Yang,
Phys. Lett. {\bf B633} (2006) 636.

\bibitem{Wu}
Y. Wu
Phys. Lett. {\bf B634} (2006) 74.

\bibitem{BMPV}
A. P. Balachandran, G. Mangano, A. Pinzul, S. Vaidya,
Int. J. Mod. Phys. {\bf A21} (2006) 3111.

\bibitem{BGMPQV}
A.P. Balachandran, T.R. Govindarajan, G. Mangano , A. Pinzul, B.A.
Qureshi, S. Vaidya,
{\bf hep-th/0608179}.



\bibitem{SGT}
T. P. Singh, S. Gutti, R. Tibrewala, {\bf gr-qc/0503116}.

\end{thebibliography}
\end{document}